\documentclass[10pt]{article}
\usepackage{epsfig}
\begin{document}

\twocolumn[\hsize\textwidth\columnwidth\hsize\csname@twocolumnfalse\endcsname

\begin{center}

{ \bf\large Absolute frequency measurement of the 435.5 nm  $^{171}$Yb$^+$ clock transition 
with a Kerr-lens mode-locked femtosecond laser}
\vspace{0.5cm}

J\"orn Stenger, Christian Tamm, Nils Haverkamp, Stefan Weyers, and Harald R. Telle
\vspace{0.25cm}

{\it Physikalisch-Technische Bundesanstalt, Bundesallee 100, 38116 Braunschweig, Germany}
\vspace{1cm}

\begin{minipage}{12cm}
We have measured the frequency of the \mbox{$6s\: ^2S_{1/2} - 5d\: ^2D_{3/2}$}
electric-quadrupole transition 
of $^{171}$Yb$^+$ with a relative uncertainty of $1\times 10^{-14}$,
\mbox{$\nu_{Yb}$ = 688 358 979 309 312 Hz $\pm$ 6 Hz. }
A femtosecond frequency comb generator was used to phase-coherently link the optical frequency 
derived from a single trapped ion to a cesium fountain controlled hydrogen maser.
This measurement is one of the most accurate measurements of optical
frequencies ever reported, and it represents a contribution to the development of optical clocks
based on an $^{171}$Yb$^+$ ion standard.
\end{minipage}
\vspace{1cm}

\end{center}

]

Frequency comb generators based on Kerr-lens mode-locked
femtosecond lasers \cite{udem99,didd00} have dramatically simplified
absolute optical frequency measurements. 
Such measurements of optical clock transitions in samples of cold atoms or single trapped ions
mark an important step towards the realization of future optical clocks.
Several optical transition frequencies have already
been directly compared with primary cesium standards by the femtosecond comb technique, 
such as the hydrogen Lyman-$\alpha$ transition \cite{nier00}, the 657~nm 
intercombination transition in Ca \cite{sten01,udem01}, or a mercury ion transition at 282~nm \cite{udem01}. 
Recently, the indium ion clock transition at 237~nm was measured with respect to a cesium-fountain calibrated 
He-Ne standard \cite{zant00}. Here we report the first absolute frequency measurement of the 
\mbox{$6s\: ^2S_{1/2}(F=0) - 5d\: ^2D_{3/2}(F=2)$} transition of $^{171}$Yb$^+$ at 435.5~nm (688~THz).

This transition is attractive for optical clocks due to its
small natural linewidth of 3.1~Hz and a \mbox{$\Delta m_F$ = 0} 
component with vanishing low-field linear Zeeman frequency shift. 
A single $^{171}$Yb$^+$ ion is laser cooled in a spherical radiofrequency Paul trap 
so that the Lamb-Dicke condition is satisfied at 435.5~nm. The clock transition is alternately probed on both 
sides of the resonance line with the frequency-doubled output of an extended-cavity 
diode laser emitting at 871.0~nm. One laser cooling and probe excitation cycle lasts 80~ms. 
The frequency of the probe laser is stabilized to the line center with an effective 
time constant of 30~s through a second-order integrating servo algorithm.
Short-time fluctuations are reduced by stabilization to an environmentally isolated 
high-finesse cavity. More details are given in Ref. \cite{tamm00}.
In the measurements reported here the $^{171}$Yb$^+$ clock transition was resolved with an essentially
Fourier-limited linewidth of 30~Hz. 

A frequency comb was generated by a Kerr-lens mode-locked femtosecond laser.
The emitted periodic pulse train corresponds in the frequency domain
to a comb-like spectrum, which can be completely characterized by only three numbers: the line
spacing equal to the repetition rate $f_{\rm rep}$, 
the longitudinal mode order $m$ of a line and an offset frequency $\nu_{\rm ceo}$, 
which reflects the frequency offset of the whole comb with respect to the frequency origin. 
Thus, an external optical frequency $\nu_{ext}$ can be written

\begin{equation}
\nu_{ext} = \nu_{ceo} + m f_{rep} +\Delta x\;,
\label{nuopt}
\end{equation}

\noindent where $\Delta x$ is the beat frequency of the external optical signal with the $m$th comb
mode. In the experiment the radiofrequencies $\nu_{ceo}$, $f_{rep}$, and $\Delta x$ are referenced
to a hydrogen maser, which in turn is compared with a cesium fountain.

\begin{figure*}[ht]
  \centerline{\includegraphics[width=12cm]{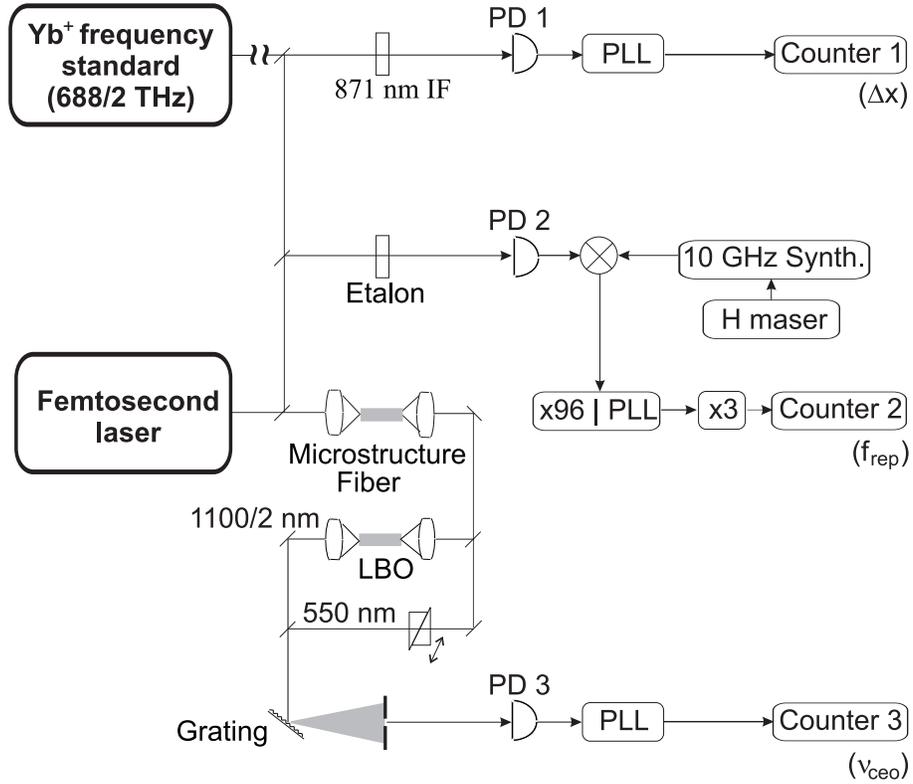}}
  \smallskip
  \caption{Schematic of the setup. LBO denotes the LiB$_3$O$_5$--frequency doubling crystal, 
PD~1-3 photo detectors,
IF interference filter, and PLL phase-locked loop, respectively. Details are described in the text.
Two additional servo loops (not shown in the Figure) were used for a slow stabilization of
$f_{rep}$ and $\Delta x$ in order to keep the beat signals within the hold-in range of the 
PLL tracking oscillators.}
\end{figure*}

The experimental setup is schematically shown in Fig.~1. 
The frequency comb generator comprises a 10~fs Kerr-lens mode-locked 
Ti:Sapphire laser and a microstructure fiber \cite{rank00} for external broadening of the spectrum via 
self-phase modulation. More details are given in Ref. \cite{sten01}. 
By coupling approximately 30~mW of the laser output into a 10~cm long piece of fiber 
we achieved a spectrum ranging from 500~nm to 1200~nm.

The 871~nm light from the $^{171}$Yb$^+$ standard was guided to the frequency comb 
generator via a 250~m long single-mode polarization preserving fiber. About 1~mW of that 
light was combined with light from the femtosecond laser. After spectral filtering by a 10~nm (FWHM)
interference filter and spatial filtering in a short piece of single mode fiber the beat note 
$\Delta x$ was detected with a fast Si PIN photodiode (PD~1), filtered by a phase-locked loop (PLL) 
and counted by a totalizing counter.

Special care had to be taken for a phase-resolved determination of the repetition rate $f_{rep}$, 
which, according to eqn. (\ref{nuopt}), enters the optical frequency measurement with a large
multiplication factor $m$.  Thus we detected the 103rd harmonic of $f_{rep}$
at 10~GHz with a fast InGaAs PIN photodiode (PD~2) after spectral filtering with a fused-silica etalon.
This microwave signal was downconverted, filtered and frequency-multiplied by 288.
Owing to the resulting large overall multiplication factor of 
29\,664 the digitization error was reduced below the instability of the hydrogen maser.
The wavelength of the 871~nm signal was pre-measured by a lambdameter with 
absolute accuracy corresponding to 1.5~MHz, thus determining the longitudinal mode order $m$.

The frequency $\nu_{ceo}$ was measured by detecting the beat note between frequency-doubled comb modes
around 1070~nm and modes around 535~nm. 
According to eqn. (\ref{nuopt}) the frequencies of the comb modes are shifted by $\nu_{ceo}$ 
whereas the harmonics are shifted by $2 \nu_{ceo}$. The resulting beat note $\nu_{ceo}$
was detected by a photo multiplier (PD~3) after spatial and spectral filtering both fields 
with a single mode fiber and a 600~l/mm grating, respectively. The signal 
was tracked with a third PLL and finally counted.

Data were taken on three different days. By averaging we derive the following value for the 
$6s\; ^2S_{1/2}(F=0) -\; 5d\; ^2D_{3/2}(F=2)$ electric-quadrupole clock transition of the $^{171}$Yb$^+$ ion:

\begin{equation}
\nu_{Yb} = 688\; 358\; 979\; 309\; 312\; {\rm Hz} \pm 6\; {\rm Hz}\;.
\label{ybfrequency}
\end{equation}

This frequency includes the frequency shift of the $^{171}$Yb$^+$ transition due to isotropic 
blackbody radiation at an ambient temperature of 298~K. This shift is calculated to $-$~0.4~Hz using 
tabulated atomic data \cite{fawc91}. 
Fig.~2 shows the Allan standard deviation of one day's data.  
The typical instability of the hydrogen maser (open circles in Fig~2) is approached. 

\begin{figure}[ht]
  \centerline{\includegraphics[width=8.2cm]{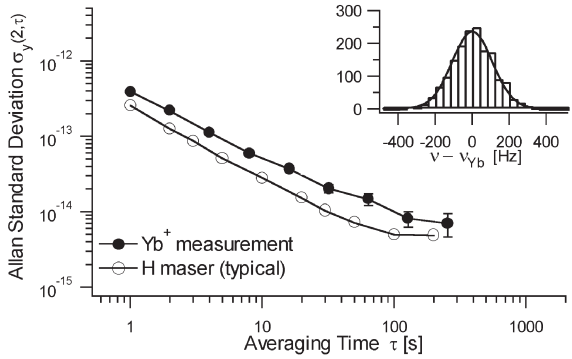}}
  \smallskip
  \caption{Allan standard deviation of the $^{171}$Yb$^+$ clock transition
measurement and that of the typical perfomance of the hydrogen maser.
The inset shows the distribution of the measured frequency values (averaging time 1~s, 
\mbox{bin-width = 40 Hz}, \mbox{$\sigma$ = 111 Hz)}. The Allan standard deviation 
of the clock transition measurement approaches the typical performance of the
hydrogen maser. The maser was operated in a self-tuning mode which reduced the available frequency 
stability for averaging times in the range of 200~s.
}
\end{figure}

The combined $1\sigma$ uncertainty of 6~Hz is given by the 
random and systematic contributions listed in Table 1. Sources of systematic uncertainties are 
the cesium fountain frequency standard \cite{weye01} and the $^{171}$Yb$^+$ frequency standard.
The contributions due to thermally and acoustically induced 
length fluctuations both of the coaxial cable carrying the 100~MHz hydrogen maser signal
and of the optical fiber guiding the 871~nm light were measured to be below $1\times 10^{-15}$ and thus are
negligible. For the $^{171}$Yb$^+$ frequency standard we take into account a servo uncertainty of 1~Hz 
due to probe laser frequency drifts which are in the range of \mbox{- 0.05 $\pm$ 0.03 Hz/s} and 
a 3~Hz uncertainty due to confinement-related shifts of the atomic transition frequency.
The dominant source of these is the electric-quadrupole interaction of the upper state of the clock transition 
with stationary electric field gradients. In order to avoid quadrupole shifts by the trap field, 
the applied trap voltage contained no dc component. The residual shift due to uncompensated 
stray field gradients is estimated to be not larger than 1~Hz for atomic 
$D_{3/2}$ and $D_{5/2}$ states \cite{wine89}. An arrangement 
of compensation coils was used to adjust a magnetic field of $1\pm 0.2 \mu$T in the trap region 
during excitation of the $^{171}$Yb$^+$ clock transition. The corresponding quadratic Zeeman shift of 
the $\Delta m_F$ = 0 reference transition is in the range of only 0.05~Hz.
The trap region was protected from ambient heat sources by the light shield of the trap setup. 
We assume that the thermal radiation field in the trap region represented an equilibrium 
blackbody field at room temperature and neglect the corresponding contribution to the uncertainty budget.

\begin{table*}[ht]
\begin{center}
\begin{tabular}{c|cc|cc|c}
\hline\hline
corrected  & \multicolumn{4}{c|}{standard uncertainties arising from:}                              & combined \\ 
 results            & \multicolumn{2}{c|}{random effects}& \multicolumn{2}{c|}{systematic effects}& uncertainty \\
$\nu - \nu_{Yb}$ &measurement & reference & \quad Yb$^+$ \quad    & reference  &         \\ \hline
day1:   +  0.5               &  12.4               & 5.4         &  3.2             & 1.0           & 13.9  \\
day2:  +  2.6               &  7.8                 & 6.3         &  3.2             & 1.8           & 10.7  \\
day3:  $-$ 2.5              &  5.6                  & 4.2         &  3.2             & 1.8           & 7.9    \\ \hline\hline
\end{tabular}
\caption{Deviations of the frequency measurement results from the weighted mean
\mbox{$\nu_{Yb} = 688\; 358\; 979\; 309\; 312 \pm 6$ Hz}, and uncertainties. 
The results are corrected for shifts of the H-maser frequency relative to the fountain reference.
All numbers are given in Hz. 
}
\end{center}
\end{table*}

In conclusion, we measured the frequency of the electric-quadrupole clock transition of the $^{171}$Yb$^+$ 
ion with a relative uncertainty of $1\times 10^{-14}$. This demonstrates the potential of 
the $^{171}$Yb$^+$ standard as an ultraprecise optical frequency reference. 
Simultaneously, we demonstrated the capability of a femtosecond comb generator of measuring 
optical frequencies with Cs clock accuracy. A future application can be the direct frequency comparison of 
the $^{171}$Yb$^+$ standard with another optical standard such as the cold-atom based calcium standard,
aiming e.g. to measure a possible variation of fundamental constants \cite{pres95}.
A drift of the relative frequencies of these optical standards 
due to a drift of the finestructure constant $\alpha$ by more than
$10^{-15}$ per year appears excluded \cite{pres95}. However, the unprecedented measurement 
accuracy achievable with femtosecond comb generators
encourages one to pursue such fundamental, ultra-precise measurements.

We gratefully acknowledge financial support from the Deutsche Forschungsgemeinschaft through 
SFB 407 and contributions of Burghard Lipphardt, Uwe Sterr, Andreas Bauch,
G\"unter Steinmeyer and Ursula Keller in several stages of the experiment. 
We are also indebted to Robert Windeler of Lucent Technologies for providing the microstructure fiber.

\end{document}